\documentclass{article}
\usepackage[utf8]{inputenc}
\usepackage{subfig}
\usepackage{graphicx}
\usepackage{listings}
\usepackage{color}
\usepackage{verbatim}
\usepackage{bm}

\title{AVATAR: Blender add-on for fast creation of 3D human models}
\author{Jordi Sanchez-Riera, Aniol Civit, Marta Altarriba, \\ and Francesc Moreno-Noguer }
\date{November 2020}

\begin{document}

\maketitle

\begin{abstract}
Create an articulated and realistic human 3D model is a complicated task, not only get a model with the right body proportions but also to the whole process of rigging the model with correct articulation points and vertices weights. Having a tool that can create such a model with just a few clicks will be very advantageous for amateurs developers to use in their projects, researchers to easily generate datasets to train neural networks and industry for game development. 
We present a software that is integrated in Blender\cite{Blender} in form of add-on \cite{avatar_addon} that allows us to design and animate a dressed 3D human models based on Makehuman\cite{Makehuman} with just a few clicks. Moreover, as it is already integrated in Blender, python scripts can be created to animate, render and further customize the current available options. 
\end{abstract}

\section{Introduction}

Human models are used in many kinds of applications. From architects that want to enhance their building renders with more natural feeling to big companies studios in games, passing through amateur developers that just want create some characters for fun. The process to create such a model is costly since typically includes not only the body design but also clothes design, texture generation and rigging.

Each one of these processes is quite complicated on their own with dedicated software for each one of the parts that require expert users trained to achieve professional results. In order to facilitate these tasks, there are several human designer suites that allows an untrained user create a professional 3D human model with just a few clicks. These software can be classified into two groups: DAZ3D\cite{daz3d}, Avastar\cite{avastar}, Character Creator \cite{character_creator}, and Makehuman\cite{Makehuman} in the first group and Manuel Bastioni Lab \cite{MBLab} and SMPL \cite{SMPL_2015} in the second group.  

In the first group, the shape of the human mesh model can be adjusted with several semantic meaning parameters, as for example: body weight, belly, breasts, etc. Moreover, a collection of clothes, different hair styles, and extra apparels that can be easily integrated into the body model. Exists also the option to customize texture patterns and colors. In the second group, the software only includes options to modify pose and shape parameters of a naked human mesh model, which in the case of SMPL model, the shape parameters don't correspond to any semantic meaning.

Typically, once the model is created it is exported to some standard file format so it can be used for a third party software in a game, in an arbitrary scene or just simply to render it. In the case we want to generate several models or change the created model, the whole process of design, export, and import, can be a bit tedious.

The ideal case would be that the human modeller can be integrated into a render software \cite{Blender, unity, unreal, autodesk} to achieve faster implementation and production timelines.  This is the case of Manuel Bastioni Lab \cite{MBLab} which can be used as an add-on for Blender software \cite{Blender}. Everything is integrated in the same environment and hence the user can adjust and render the human model without changing the production environment. However, in Manuel Bastioni Lab, clothes, hairs or other kind of extras are not provided.

Makehuman, in the latest version, made an effort towards the integration with other software providing a communication protocol so both applications can message each other and obtain some kind of integration. Unfortunately, this kind of messaging is not very practical and it is necessary to install extra libraries in the OS as well as inside the Blender software in order to make the messaging possible.

We design a new software based on Makehuman in a form of a Blender add-on. This add-on \cite{avatar_addon}, can easy create a human body model that can accelerate the production timeline for any kind of project. The main characteristics of this add-on are: 
\begin{itemize}
    \item Has several semantic meaning parameters to adjust body shape. 
    \item It allows to load motion files to the body skeleton.
    \item Has a library of clothes that can be easily integrated to the body model
    \item All the options can be modified by a python script
    \item User can add more clothes, change textures, to the clothes library
    \item Can work with a MoCap add-on \cite{Sanchez-Riera-mocap}
\end{itemize}

\section{Avatar add-on}

\begin{figure}[t]
\centering
\subfloat[Body Panel]{\includegraphics[width=0.18\textwidth]{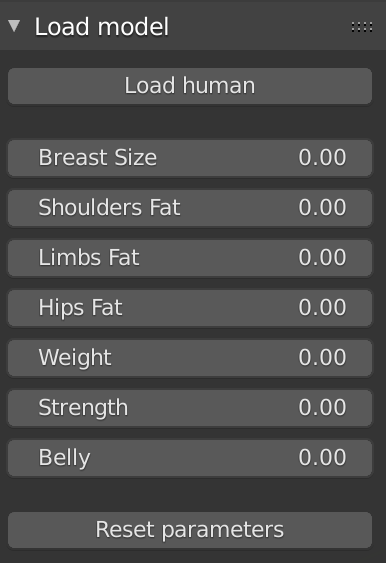}}
\subfloat[Dress Panel]{\includegraphics[width=0.49\textwidth]{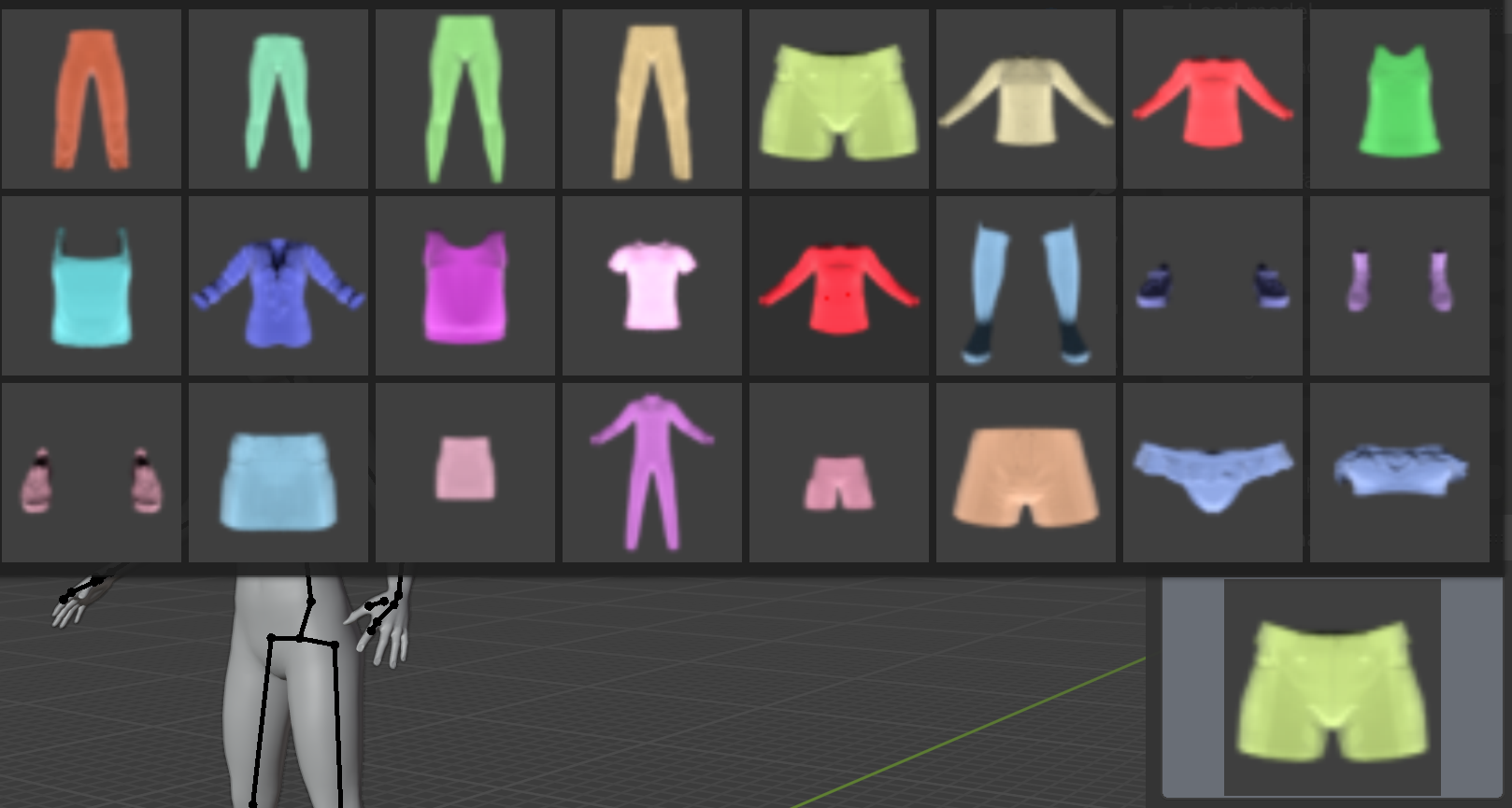}}
\subfloat[Motion Panel]{\includegraphics[width=0.322\textwidth]{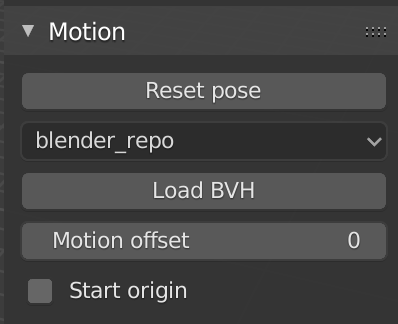}}
\caption{\textbf{Avatar add-on panels.} It displays the three panels at the beginning where the user can control easily some of the options of the add-on}
\label{fig:addon_panels}
\end{figure}

The addon is divided into several panels, as shown in Figure \ref{fig:addon_panels}. One panel controls the shape of the body, other panel is a library of clothes that one could easily add to the character and could be extended to hair or other properties. And finally there is a panel to transfer motion from BVH files to the character. The Makehuman model is also compatible with Motion character \cite{retarget_bvh}.

\subsection{Shape Panel}

Makehuman has many parameters to adjust the body model. The number of parameters is too big, you can model the face shape (round, oval, squared, etc.), some details on face (nose, ears, eyes) and similar for the body. This is mostly intractable and not very useful when want to design a fast human prototype.
Therefore, we will define our mesh model $\mathbf{x} = [v_i, .., v_n]$ with $n$ vertices $v_i$ as a linear combination of the body rest pose $x_r$ with a set of $m$ attributes $\mathbf{A} = [A_i, .., A_m]$
\begin{equation}
\mathbf{x} = \mathbf{x}_r + \mathbf{A} \, \bm{\alpha} = \mathbf{x}_r + \sum_{i=1}^m A_i \alpha_i  
\end{equation}
where $\bm{\alpha} = [\alpha_i, .., \alpha_m]$ are the weights that control each one of the attributes. Typically, these attributes are obtained by applying Principal Component Analysis (PCA) over a sample of human meshes with different shapes. However, when performing PCA over the whole set of meshes, these attributes lose their semantic meaning. In order to keep the semantic meaning, the PCA is performed over subsets that only modify the body shape for a single semantic meaning, in concrete, we define $m = 7$ semantic attributes as shown in Figure\ref{fig:addon_panels}(a).

\subsection{Dress Panel}

Many contributors have designed all kind of clothes, hairs, and extra object meshes that are used in a daily basis to create a realistic human model. Unfortunately, most of these meshes have very poor textures and when applied to the model the appearance is not very realistic. Therefore, we select a subset of clothes to include in the Avatar add-on that contain high resolution textures along with normal maps to obtain a high quality human model.
When animating the Makehuman model, the skeleton controls the body and cloth vertex weights of a certain pose, and this often provokes that the vertices of the body get outside the cloth generating an incorrect geometry and texture. For this reason, the vertices of the human body mesh that overlap with the piece of cloth are removed. 
However, this is not the behavior of the Avatar add-on since we want to keep both body and cloth meshes to calculate collisions in case a cloth physics simulation is needed. Thus, each cloth mesh needs is slightly modified to ensure all vertices are outside the body mesh.

\subsection{Motion Panel}

There are basically two ways of animating a character: one is to define for each joint of our skeleton model the position at each time instance; the other one is to transfer to our character a previously recorded motion with a MoCap system. The first one requires of many time and good skills on animation, while the second one is faster and can be automatized using motion files.
One of the formats to store these motion files is the Biovision Hierarchy (BVH). This format consists into a section that describes the object skeleton and another section that describes the rotations and locations of each joint described in the first section.
The skeletons described in the BVH files are unlikely to match the skeleton of the target model, not only in size and number of joints, but also in the axis reference definition for each joint. Thus, before transfer any motion from the BVH file (source) to our human model (target) we need to do the following steps:
\begin{enumerate}
    \item Scale source skeleton to match with target skeleton
    \item Find bone correspondences between source and target skeletons
    \item Find local rotation matrices to get from source to target axis reference system
    \item Transfer rotations according to the BVH file
\end{enumerate}

\section{Scripting}

One of the major advantages of having a complete integrated environment for a character development is the capacity of create scripts, and binaries that can run as standalone application, thus, having the capacity to create dynamic and interactive environments.
This advantage has already been exploited by the research community where there is a need to obtain annotated data in order to train neural networks algorithms. These scripts allows to automatize many variations on shape, textures and poses of different models that otherwise would be nearly impossible to obtain with real data. Not only for the difficulties to find enough variety of people, but also for the difficulties to manually annotate all the possible 3D data. 
An example of such dataset is the SURREAL dataset \cite{varol17_surreal}, which contains around $2M$ images with $4$ camera viewpoints, different poses, and textures projected onto a naked body. Moreover, the dataset contains extra information such as depth, segmentation of body parts, segmentation of clothes and normals.  Similarly, the 3DPeople dataset \cite{pumarola20193dpeople} is an improved version with an actual geometry for clothes, as a correct representation for the body geometry is necessary to obtain accurate 3D body pose reconstructions. 
With the new Avatar add-on, we provide the capacity to create such datasets or new ones depending on the needs of each one of the researchers \cite{SIZER_Dataset, bertiche2020cloth3d, zhu2020deep}. 

\section{Examples}

In Figure\ref{fig:examples} we illustrate some of the possible configurations that can be done with the Avatar add-on. In the first row is shown some of the shape variations, while in the other two rows is shown some of the available clothes and the character performing some actions while wearing the clothes.

\section{Source code and other resources.}

The source code of the Avatar add-on as well as the texture images, and shape weights to control the parametric human model can be found in:
\begin{verbatim}
https://github.com/jsan3386/avatar
\end{verbatim}

\begin{figure}
\centering
\includegraphics[width=0.3\textwidth]{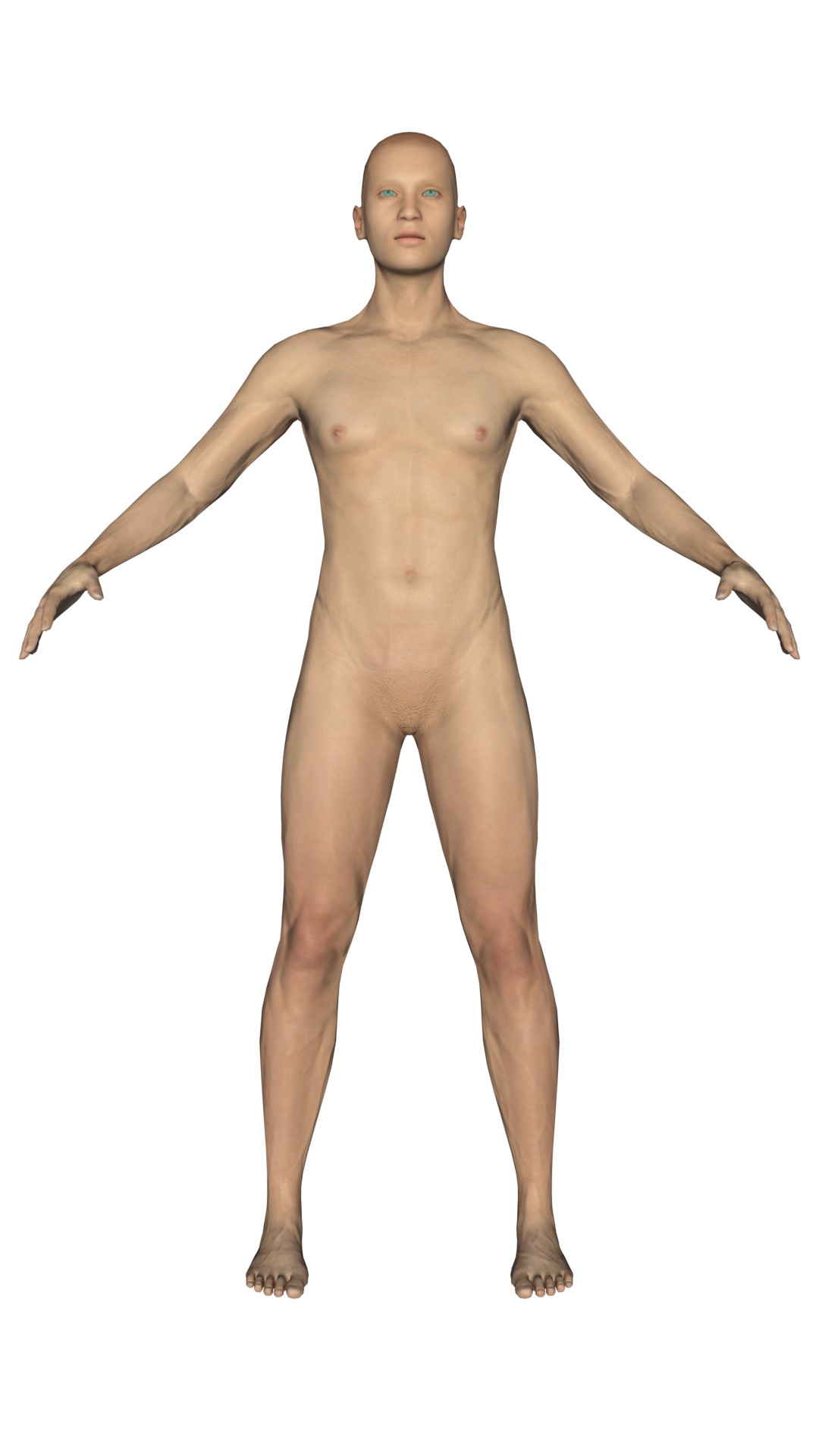}
\includegraphics[width=0.3\textwidth]{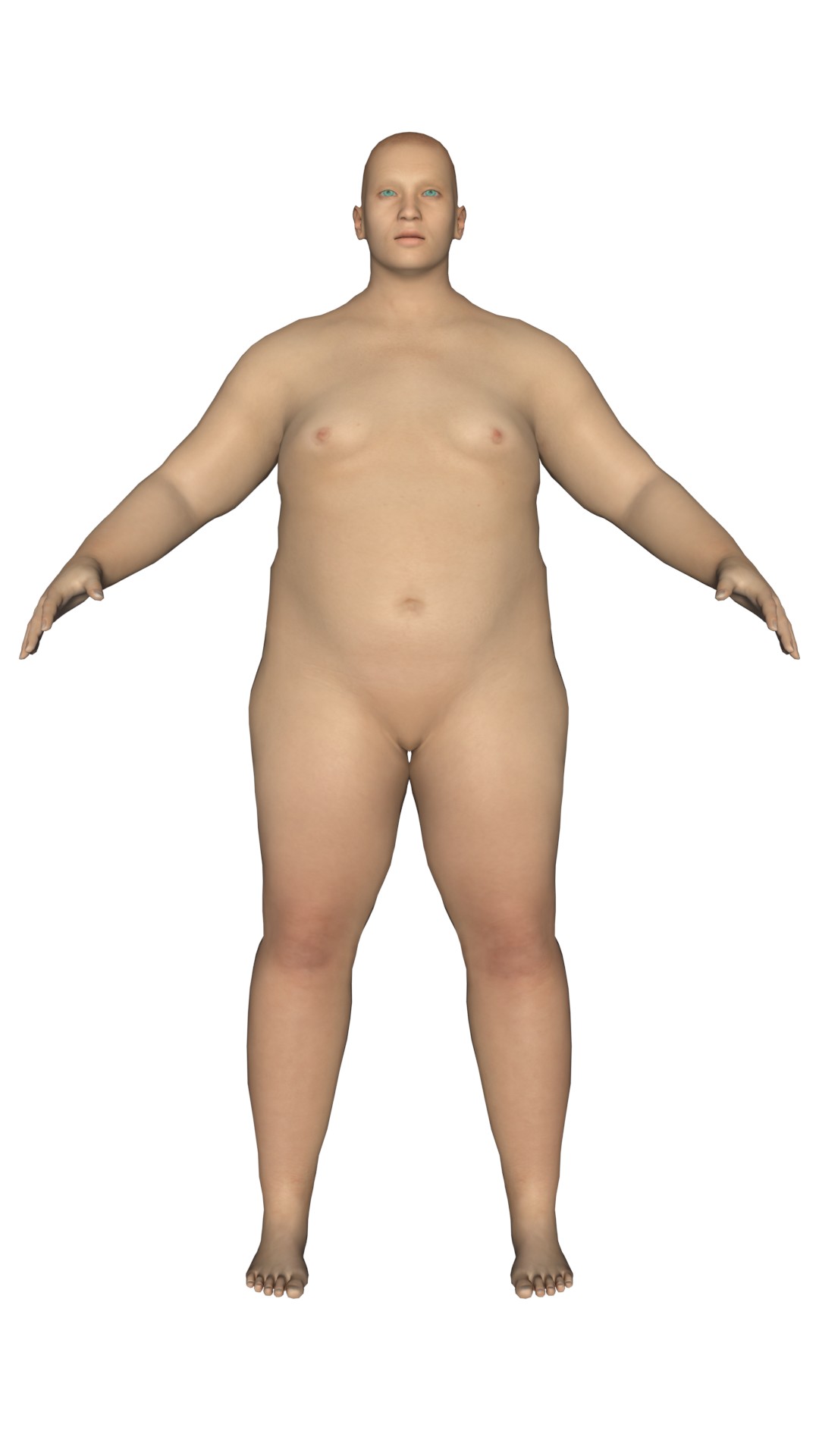}
\includegraphics[width=0.3\textwidth]{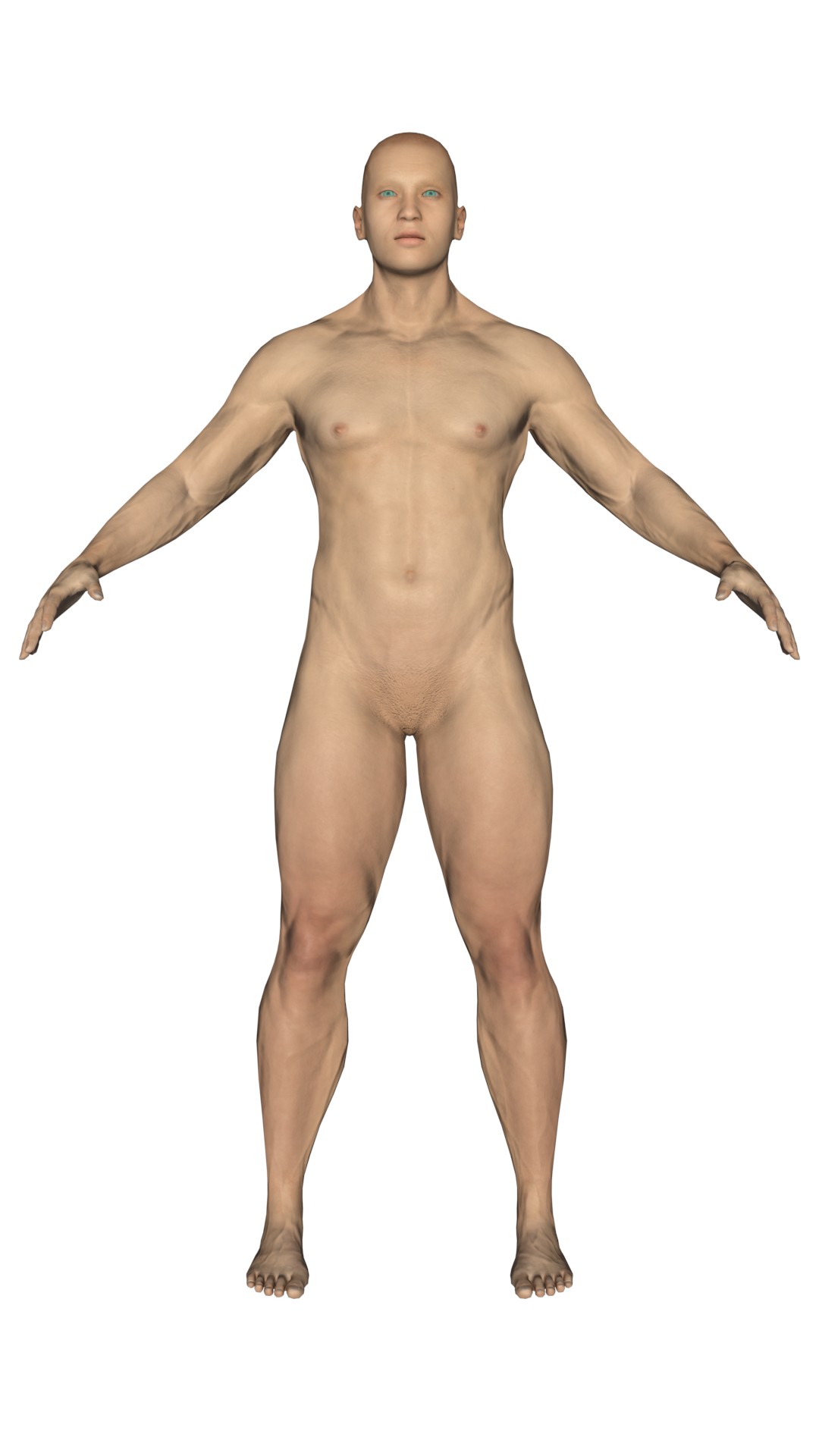} \\

\includegraphics[width=0.3\textwidth]{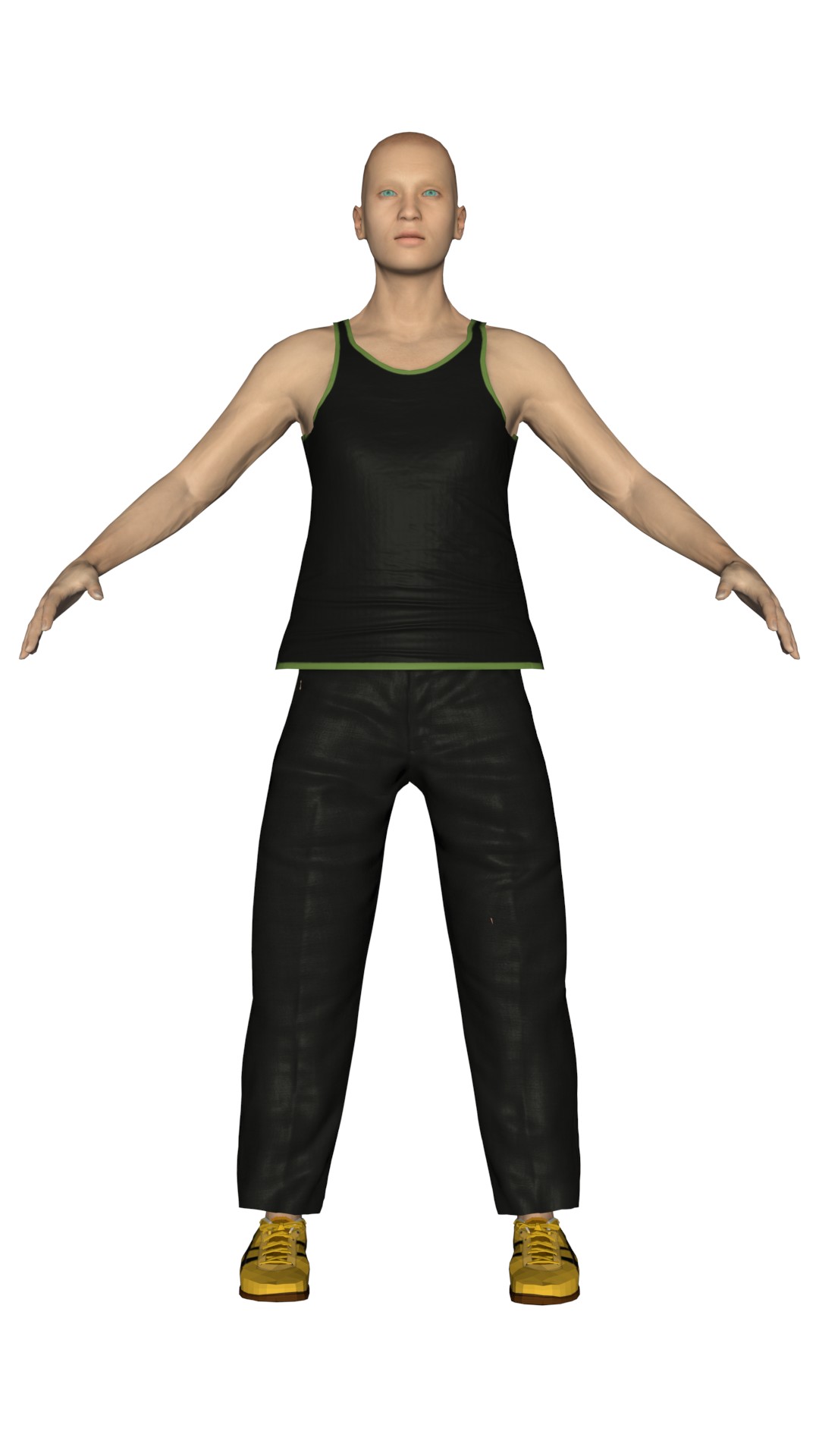}
\includegraphics[width=0.3\textwidth]{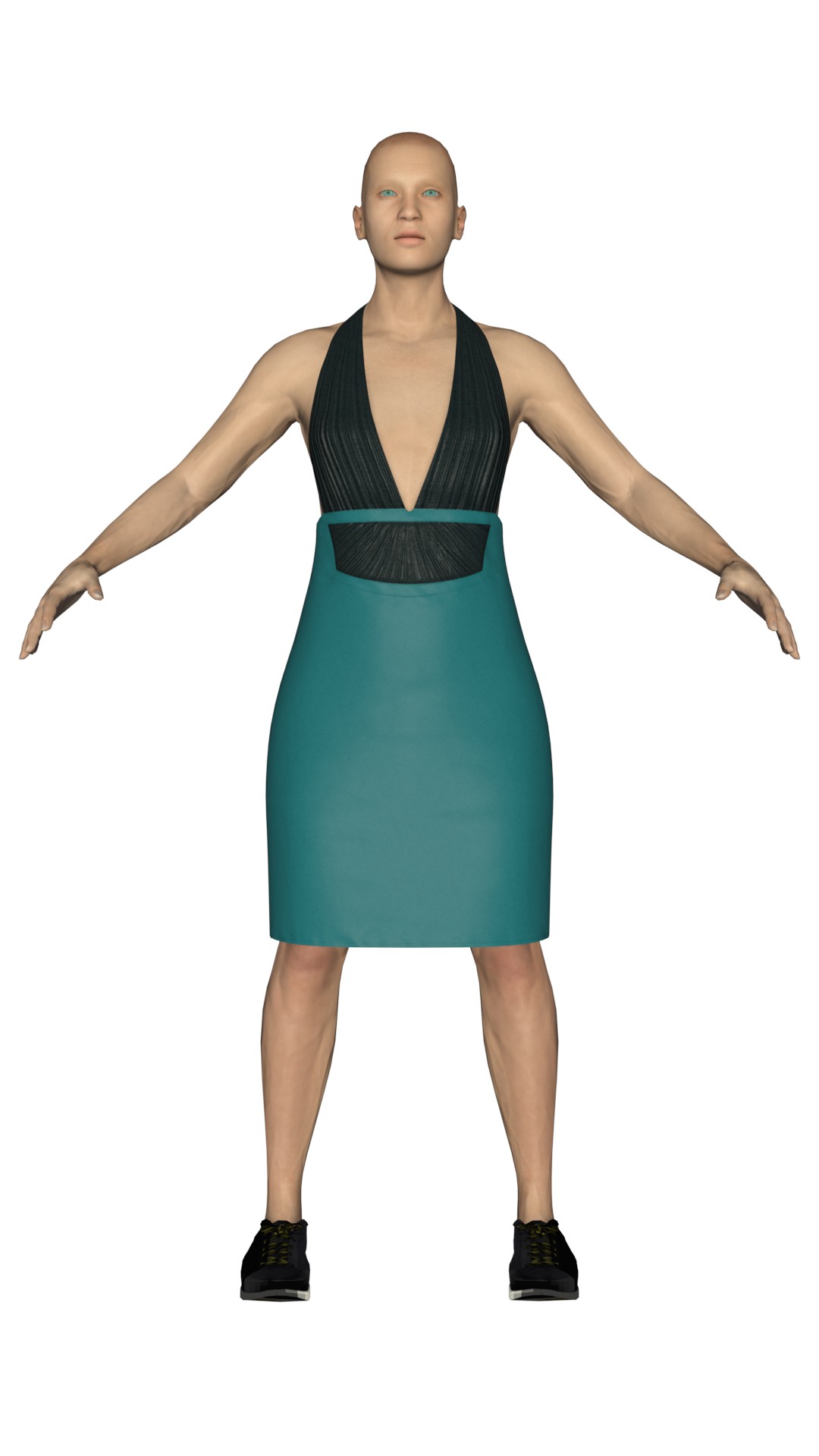}
\includegraphics[width=0.3\textwidth]{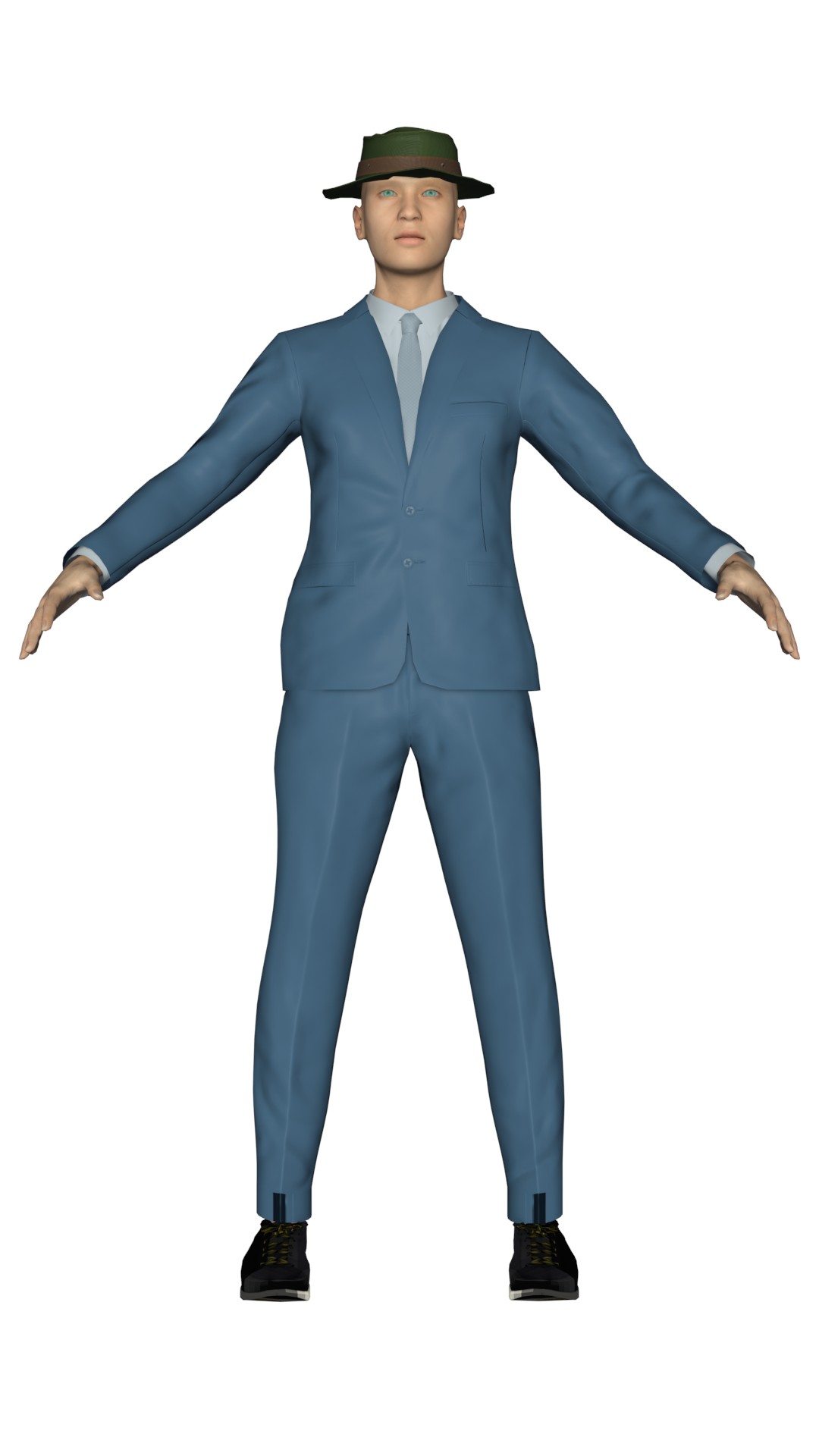} \\

\includegraphics[width=0.3\textwidth]{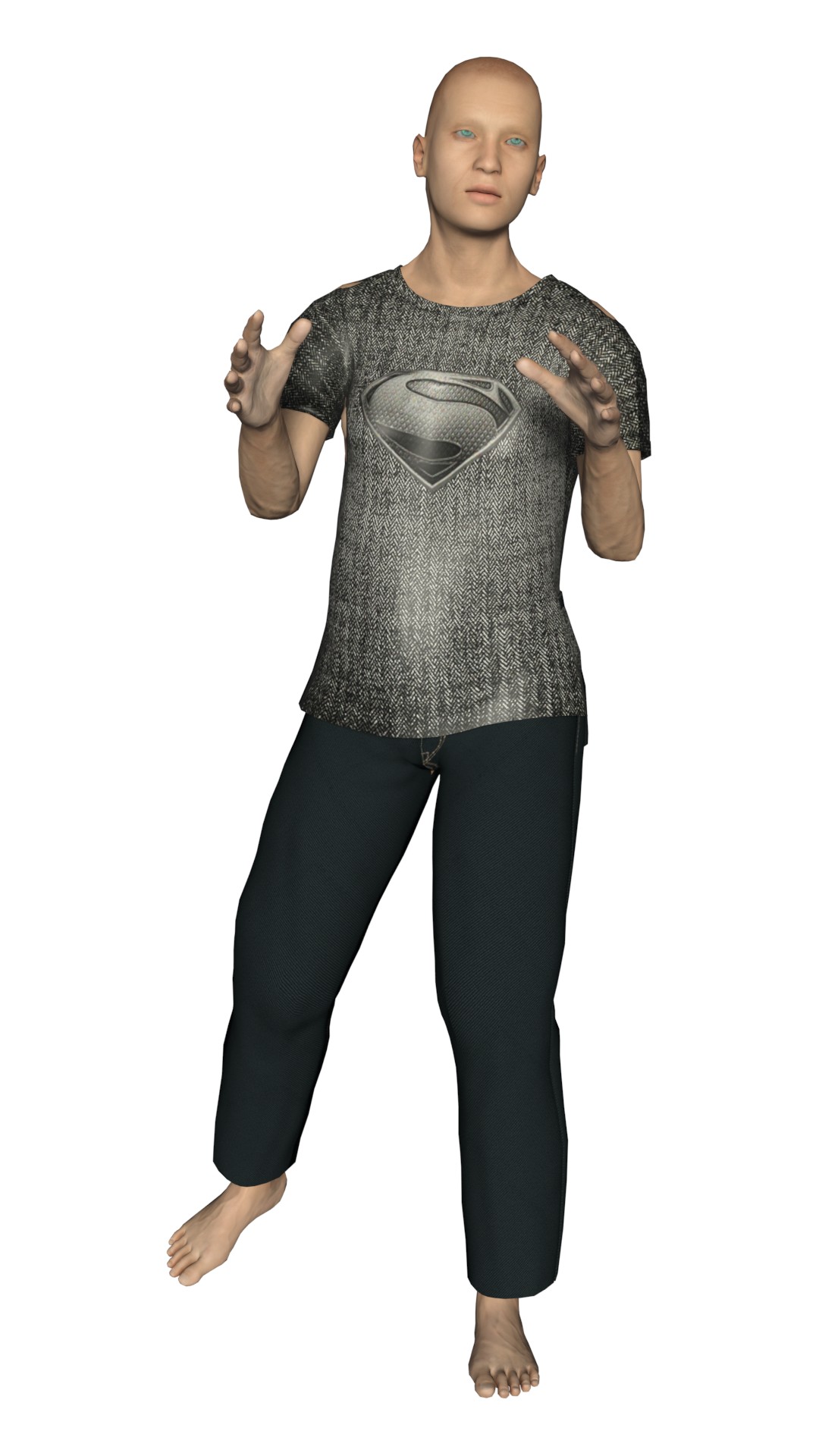}
\includegraphics[width=0.3\textwidth]{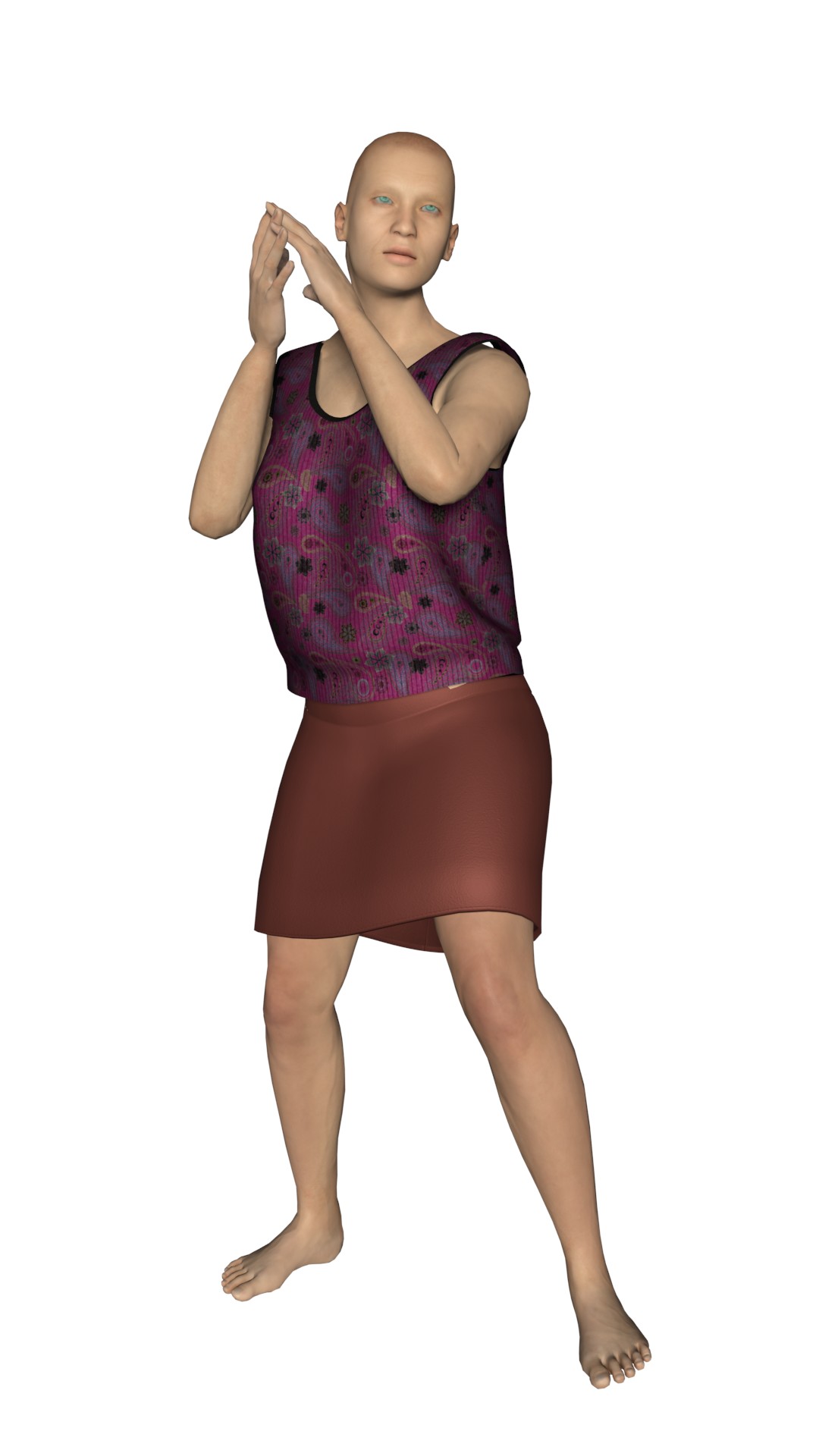}
\includegraphics[width=0.3\textwidth]{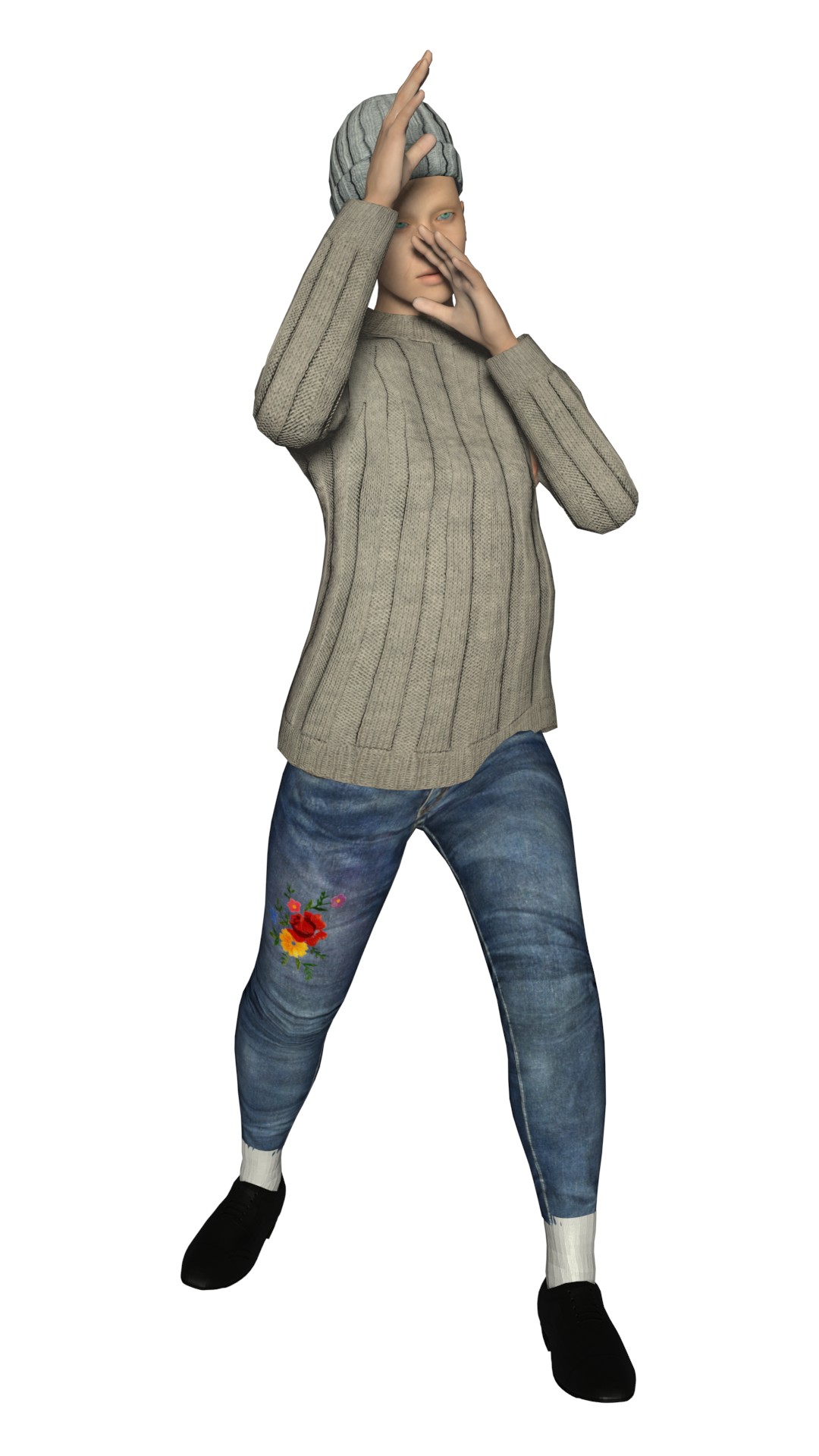}

\caption{Example of different shapes, poses and clothes that can be used in the Avatar add-on.}
\label{fig:examples}
\end{figure}

\bibliographystyle{unsrt}
\bibliography{refs}

\end{document}